# Ultrafast Generation of Pseudo-magnetic Field for Valley Excitons in WSe2 Monolayers

Jonghwan Kim*[1], Xiaoping Hong*[1], Chenhao Jin[1], Su-Fei Shi[1,2], Chih-Yuan S. Chang[3], Ming-Hui Chiu[4], Lain-Jong Li[3,4], Feng Wang[1,2,5]

[1] Department of Physics, University of California at Berkeley, Berkeley, CA 94720, United States.

[2] Material Science Division, Lawrence Berkeley National Laboratory, Berkeley, CA, 94720, United States.

[3] Institute of Atomic and Molecular Sciences, Academia Sinica, Taipei 10617, Taiwan.

[4] Physical Sciences and Engineering, King Abdullah University of Science and Technology, Thuwal, Saudi Arabia.

[5] Kavli Energy NanoSciences Institute at University of California Berkeley and Lawrence Berkeley National Laboratory, Berkeley, CA 94720, United States.

* These authors contribute equally to this work

Correspondence to: fengwang76@berkeley.edu

A new degree of freedom, the valley pseudospin, emerges in atomically thin two-dimensional transition metal dichalcogenides ($MX_2$) and has attracted great scientific interest[1,2]. The capability to manipulate the valley pseudospin, in analogy to the control of spin in spintronics, can open up exciting opportunities in valleytronics. Here we demonstrate that an ultrafast and ultrahigh valley pseudomagnetic field can be generated using circularly polarized femtosecond pulses to selectively control the valley degree of freedom in monolayer $MX_2$. Employing ultrafast pump-probe spectroscopy, we observed a pure and valley-selective optical Stark effect in $WSe_2$ monolayers from the non-resonant pump, which instantaneously lift the degeneracy of valley exciton transitions without any dissipation. The strength of the optical Stark effect scales linearly with both the pump intensity and the inverse of pump detuning. An energy splitting more than 10 meV between the K and K' valley transitions can be achieved, which corresponds to an effective pseudomagnetic field over 170 Tesla. Our study demonstrates efficient and ultrafast control of the valley excitons with optical light, and opens up the possibility to coherent manipulate the valley polarization for quantum information applications.

Atomically thin layers of transition metal dichalcogenides ($MX_2$) have emerged as an exciting two-dimensional semiconductor platform for nanoelectronics and optoelectronics[1,3]. In particular, a pair of degenerate bands are present at the K and K' valleys in the momentum space of hexagonal $MX_2$ monolayers, giving rise to a unique valley degree of freedom that is analogous to electron spin[2]. Recent polarization-resolved photoluminescence (PL) studies show that the valley pseudospin in $MX_2$ can couple directly to the helicity of excitation photons[4-7], and the pseudospin polarization between two valleys exhibits a remarkable coherent behavior[7]. It raises the intriguing prospect of valleytronics that exploits the valley degree of freedom to carry information[1,2,4-9].

Just as spin manipulation in spintronics, the capability to control the valley pseudospin is essential for valleytronics based on $MX_2$ materials. In spintronics, the electron spin can be manipulated through any external perturbation that breaks the energy degeneracy of two orthogonal spin polarizations. This can be achieved either through an external magnetic field, the most common approach[10,11], or through a pseudomagnetic field generated by other stimuli. For example, circularly polarized light can produce a pseudomagnetic field and rotate the electron spin through the optical Stark effect. (The optical Stark effect, a well-established phenomenon in atomic physics and quantum optics, describes the energy shift in a two-level system induced by a non-resonant laser field.) Such pseudomagnetic field generated by spin-selective optical Stark effects has been demonstrated as an effective and ultrafast way to manipulate electron spins in semiconductor quantum wells and quantum dots[12,13]. It will be highly desirable to realize similar control of valley excitons in $MX_2$ using light-induced pseudomagnetic field for valleytronics.

In this report, we demonstrate a valley-selective optical Stark effect in monolayer $WSe_2$ to generate ultrafast and ultrahigh pseudomagnetic field for valley excitons. Previously, resonant pump-probe spectroscopy has been used to probe ultrafast carrier dynamics after the pump photons are absorbed by exciton transitions in MX$_2$ monolayers[14-17]. Here we employ non-resonant circularly polarized laser pumping with photon energies below the bandgap, which induces a coherent and dissipationless valley-selective optical Stark effect for pseudospin manipulation. We show that transition energies of the degenerate exciton resonances at K and K' valleys can be selectively shifted using circularly polarized laser pulses. The optical Stark energy shift happens instantaneously at the presence of a pump pulse, and its magnitude shows systematic dependence on the pump intensity and detuning. The light-induced valley exciton energy splitting can be as large as 10 meV, which corresponds to a pseudomagnetic field ~170 Tesla. Such valley-selective optical Stark shift will be enabling for ultrafast and all-optical control of the valley polarization in atomically thin MX$_2$ layers.

We used triangular monolayers of WSe$_2$ grown by chemical vapor deposition on sapphire substrates[18]. Optical micrograph of a typical monolayer WSe$_2$ sample is displayed in the inset of Fig. 1a. The optical reflection spectrum of monolayer WSe$_2$ at 77 K exhibits two prominent resonances at 1.68 eV and 2.1 eV (Fig. 1a), which correspond, respectively, to the A- and B-excitons split by the spin-orbital coupling[19,20]. In this study, we focused on the lowest energy A-exciton transition. This A-exciton is well separated from other excited states in energy due to strong electron-hole interactions in monolayer WSe$_2$. As a result, we can approximate the ground state and the A-exciton as a two-level system and neglect the effect from other excited states. In monolayer WSe$_2$ the A-exciton states at K and K' valleys are time-reversal pairs and have degenerate energy levels. However, they have distinctive optical selection rules and couple to photons of opposite helicity [2,4]. The dipole transition matrix element of the K-valley A-exciton is characterized by $P_K^{\sigma\pm} = \langle K|\hat{p}_x \pm i\hat{p}_y|0\rangle$, where $|K\rangle$ and $|0\rangle$ correspond to the K-valley A-exciton and ground state, respectively, $\hat{p}_x, \hat{p}_y$ are momentum operators, and $\sigma_+ / \sigma_-$ corresponds to left/right circular light. In MX$_2$ monolayers $P_K^{\sigma+}$ has a finite value but $P_K^{\sigma-}$ is approximately zero. Consequently, the A-exciton at the K valley couples exclusively to left-circularly polarized light. The A-exciton at the K' valley, on the other hand, couples only to the right-circularly polarized light. This dipole selection rule for valley transitions in monolayer WSe$_2$ governs both resonant and non-resonant excitations, which can have different manifestations in valley physics.

The valley-dependent resonant excitation enables both the generation of a valley-polarized exciton population through the absorption of circularly polarized light and the detection of valley-polarized excitons through polarization-resolved photoluminescence (PL)[4-6]. Figure 1b displays polarization-resolved PL spectra of our WSe$_2$ samples upon laser excitation at 1.8 eV with $\sigma_+$ circular polarization, where the PL intensity with helicity matching the excitation light ($\sigma_+$, red curve) is about four times stronger than that for the opposite helicity ($\sigma_-$, blue curve). This behavior is similar to that reported in previous studies[4-7], and verifies the valley-selective excitation and emission in our WSe$_2$ monolayers. The non-zero $\sigma_-$ polarized PL is presumably due to a finite intervalley scattering because the excitation energy is ~120 meV higher than the A-exciton resonance.

To manipulate the valley polarization, however, non-resonant coupling based on the optical Stark effect is more advantageous because it avoids the dissipation and dephasing naturally accompanying real excitations. The optical Stark effect is widely used in quantum optics to control a rich variety of quantum systems, ranging from the manipulation of cold atoms [21] and individual trapped ions[22] to coherent control of superconducting qubits[23] and electron spins [24]. In a two-level system, the optical Stark effect can be readily understood using the dressed atom picture[25]. Consider the two-level system composed of the ground state and A-exciton state at the K (or K') valley in a monolayer WSe$_2$. Figure 2a illustrates the effect of a left circularly polarized ($\sigma_+$) pump with photon energies below the exciton resonance (red arrow). In the dressed atom picture, the dressed ground state (with N $\sigma_+$ photons) and the K-valley A-exciton (with N-1 $\sigma_+$ photons) are coupled by the dipole transition, which leads to wavefunction hybridization and energy level repulsion. It effectively shifts down the ground state energy and shifts up the K-valley exciton energy. The A-exciton at the K' valley, on the other hand, cannot couple to the ground state with an extra $\sigma_+$ photon due to the optical selection rule, and the related states are not shifted by $\sigma_+$ polarized light. As a result, the energy degeneracy between K and K' valley states is lifted by the valley-selective optical Stark effect, which can be characterized by a valley pseudomagnetic field.

We measured the valley-selective optical Stark shift in WSe$_2$ monolayers with non-resonant circularly polarized excitation using pump-probe spectroscopy. Femtosecond pump pulses with tunable photon energies below the A-exciton resonance were generated by an optical parametric amplifier. Optical Stark shifts of the WSe$_2$ exciton transitions induced by the pump pulses was then probed in transient reflection spectra over the spectral range of 1.59-1.77 eV using a laser-generated supercontinuum light. Figure 2b and 2c display two-dimensional plots of transient reflection spectra in a monolayer WSe$_2$ with $\sigma_+$ and $\sigma_-$ polarized probe light, respectively, upon $\sigma_+$ polarized pump excitation. Here the non-resonant pump photons are at 1.53 eV, which is 150 meV below the exciton resonance and do not excite any real transitions. The colour scale in Fig. 2b and 2c represents the relative reflectivity change ΔR/R, the horizontal axis shows the probe photon energy, and the vertical axis shows the pump-probe time delay. For atomically thin WSe$_2$ layers on a transparent sapphire substrate, the reflection change *ΔR/R* is directly proportional to the change in the absorption coefficient[26,27]. It is apparent that strong changes in the exciton absorption are present only for $\sigma_+$ probe pulses (Fig. 2b), and no pump-induced signals can be detected above the noise level for $\sigma_-$ probe pulses (Fig. 2c). This indicates that the non-resonant $\sigma_+$ pump significantly modifies the A-exciton transition at the K valley, but not at the K' valley.

To better examine the time evolution and spectral lineshape of the photo-induced transient reflection spectra in WSe$_2$ monolayers, we plot in Fig. 3 vertical and horizontal linecuts of the two-dimensional plot for the $\sigma_+$ pump and $\sigma_+$ probe configuration (Fig. 2b). Figure 3a displays the transient reflection signals at the probe photon energy below (1.65 eV, blue line) and above (1.7 eV, red line) the exciton resonance at different pump-probe delay $\tau$ (vertical linecuts in Fig. 2b). Both signals have the largest magnitude at $\tau=0$ and show a symmetric rise and decay dynamics. This is characteristic of an instantaneous response with the rise/decay time limited by the laser pulse width of ~ 250 fs. It shows that the exciton absorption is modulated only when the pump radiation is present. In addition, the transient reflectivity has opposite signs for probe energies at 1.65eV and 1.7 eV. Detailed transient reflection spectrum $\Delta R/R$, which is proportional to the transient absorption change[26,27], is shown in Fig. 3b (green circles) for $\tau=0$ through a horizontal linecut in Fig. 2b. We observe that the transient absorption signal changes sign at exactly at the A-exciton resonance energy $E_A=1.68$ eV. The pump laser leads to reduced absorption below $E_A$ and increased absorption above $E_A$, which matches well the linear derivative of the WSe$_2$ monolayer absorption spectrum (magenta line in Fig. 3b). Quantitatively, the photo-induced absorption change can be perfectly described by a simple blueshift of ~ 4 meV of the K-valley exciton resonance. The blueshift in transition energy, instantaneous time response, and valley selectivity demonstrate unambiguously the optical Stark effect in monolayer WSe$_2$ upon non-resonant pump excitation. We note that the photo-induced signal becomes non-detectable immediately after the pump pulses, indicating negligible dissipative processes associated with real exciton absorption. We also performed the pump-probe spectroscopy using linearly polarized pump and probe pulses

(see Supplementary Information Part 1). Here photo-induced absorption changes are of similar magnitude independent of the pump and probe polarizations. This is because the linear-polarized pump has both $\sigma_+$ and $\sigma_-$ components, which lead to optical Stark shifts for both K and K' valley exciton transitions.

Next we examine how the valley-selective optical Stark shift varies with the pump laser intensity and detuning. The pump detuning $\hbar\Omega$ is defined as the difference between the exciton resonance energy energy ($\hbar\omega_0$) and the pump photon energy ($\hbar\omega_p$). Figure 4 shows the transient reflection data (left axis) and the corresponding energy shift of the A-exciton energy (right axis, see Supplementary Information Part 2) at the K valley for $\sigma_+$ pump and $\sigma_+$ probe configuration with different pump detuning (green dots in Fig. 4a) and pump intensities (green dots in Fig. 4a). We observe that the optical Stark shift is inversely proportional to the pump detuning (magenta dashed line in Fig. 4a), and scales linearly with the laser intensity (magenta dashed line in Fig. 4b). Such scaling matches well with the theoretical prediction of optical Stark shift $\delta(\hbar\omega_0) = 2S \cdot E_p^2/\hbar\Omega$, where S is the optical Stark shift constant related to the transition dipole moment and $E_p$ is the electric field of the pump pulse[25,28]. From the experimental data, we can determine an optical Stark shift constant S ~ 45 Debye$^2$ for A-exciton in monolayer WSe$_2$, which is of similar magnitude to that for exciton transition in semiconductor quantum wells[29].

The valley-selective optical Stark shift breaks the degeneracy of valley exciton transitions in monolayer WSe$_2$ and defines an effective valley pseudomagnetic field. In our experiment, the photo-induced energy splitting between K and K' exciton transitions can be as large as 10 meV (Fig. 4 and Supplementary Information Part 2). The corresponding pseudomagnetic field B$_{eff}$ for valley excitons can be estimated by $B_{eff} = \frac{\Delta E}{2g_{ex}\mu_B}$, where $\mu_B$ is the Bohr magneton and $g_{ex}$ is the effective g-factor for valley exciton transitions in WSe$_2$. The effective exciton g-factor $g_{ex}$ combines contributions from both electrons and holes, and has a theoretically predicted value ~ 0.5 for WSe$_2$ (see Supplementary Information Part 3). Using this g-factor, we estimate a pseudomagnetic field $B_{eff}$ as high as 170 T for a 10 meV splitting of valley exciton transitions. A real magnetic field of this magnitude is beyond the capability of current magnet technology, but such a pseudomagnetic field for MX$_2$ valley excitons can be produced conveniently and with femtosecond temporal control using light pulses.

In summary, we demonstrate that the optical Stark effect can generate ultrafast and ultrahigh pseudomagnetic field for valley excitons in MX$_2$ layers. It has been reported recently that excitons in different valleys in monolayer WSe$_2$, resonantly excited by linear polarization light, can maintain their phase coherence over extended time[7]. The valley-dependent optical Stark effect offers a convenient and ultrafast way to lift the valley degeneracy and introduce a controlled phase difference between the two valley states, therefore enabling coherent rotation the valley polarization with high fidelity. In analogy with spintronics, such coherent manipulation of valley polarization can open up fascinating opportunities for valleytronics.

**Methods**

**WSe$_2$ monolayer Growth:** The WSe$_2$ monolayer flakes on sapphire substrates were synthesized via chemical vapor deposition utilizing sublimed WO$_3$ and selenium powders in a horizontal tube furnace[18]. The WO$_3$ was placed at the center of the tube furnace and heated to 925°C and the selenium was placed at the upstream side of the tube furnace and maintained at 290°C *via* a heat tape. Selenium vapor was transported to the hot zone with the gas mixture 60 sccm of Ar and 6 sccm of $H_2$, while the pressure was maintained at 7 Torr during the growth. The triangular $WSe_2$ flakes were grown on the sapphire substrates located at the downstream side from $WO_3$.

**Photoluminescence:** We used a 690 nm laser (photon energy = 1.8 eV) to excite a $WSe_2$ monolayer. The laser beam was focused to a diffraction-limited spot with a diameter ~ 1 μm, and the PL was collected in the reflection geometry using a confocal microscope. A monochromator and a liquid-nitrogen cooled CCD were used to record the PL spectra.

**Linear absorption spectra:** A supercontinuum laser (Fianium SC450) was used as broadband light source. The laser was focused at the sample with ~2 μm beam size and the reflection signal R was collected and analyzed by a spectrometer equipped with a one-dimensional CCD array. The reference spectrum $R_0$ was taken on the sapphire substrate nearby the $WSe_2$ monolayer sample. The normalized difference signal $(R-R_0)/R_0$ is directly proportional to the linear absorption from atomically thin layers on sapphire.

**Pump-probe measurement:** Femtosecond pulses at 1026 nm were generated by a regenerative amplifier seeded by a mode-locked oscillator (Light Conversion PHAROS). The femtosecond pulses (at a repetition rate of 150 kHz and a pulse duration ~250 fs) were split into two parts. One part was used to pump an optical parametric amplifier to generate tunable excitation laser pulses, and the other part was focused into a sapphire crystal to generate a supercontinuum white light (500 nm ~ 900 nm) for probe pulses. The pump and probe beams were focused at the sample with a diameter of 16 μm and 8 μm, respectively. The probe light was detected by a high-sensitivity photomultiplier after wavelength selection through a monochrometer with spectral resolution of 1 nm. The pump-probe time delay was controlled by a motorized delay stage, and the pump-probe signal was recorded using the lock-in detection with a chopping frequency of 1.6 kHz. The polarization of pump and probe pulses was independently controlled using broadband quarter-wave plates and linear polarizers.

**Figure captions**

**Figure 1: Valley exciton transitions in monolayer WSe₂. a.** The optical reflection spectrum of a WSe$_2$ monolayer on a sapphire substrate at 77 K, which exhibit strong A- and B-exciton resonances at 1.68 eV and 2.1 eV, respectively. The inset shows optical microscopy image of the sample. The scale bar corresponds to 25 µm. **b.** Polarization-resolved PL spectra of a WSe$_2$ monolayer at 77K. For 1.8 eV excitation laser with left circular polarization, the PL spectra show a prominent emission peak at the A-exciton resonance (1.68 eV), and the PL intensity with left circular polarization (red curve) is about four times stronger than that with right circular polarization (blue curve). It demonstrates that valley-polarized A-exciton population can be created by circularly polarized resonant excitation, and it can be detected by polarization-resolved PL spectroscopy.

**Figure. 2: Valley-dependent optical Stark effect with non-resonant circularly polarized pump. a.** Schematic illustration of the optical Stark effect for valley transitions with non-resonant, left circularly polarized $\sigma_+$ pump. The dashed black and yellow lines denote the unperturbed ground and exciton states, respectively, and the pump photon energy is lower than the exciton resonance energy (red arrow). In the dressed atom picture, the dressed ground state (with N $\sigma_+$ photons) and the K-valley A-exciton (with N-1 $\sigma_+$ photons) are coupled by dipole transition, which results in energy level repulsion (solid black and yellow lines on the left section) and an increased exciton transition energy. On the other hand, $\sigma_+$ pump does not affect K'-valley A-exciton resonance due to optical selection rule (right section). **b. and c.** Transient reflection spectra of A-exciton resonance at 77 K. The colour scale, horizontal axis, and vertical axis represent the relative reflectivity change ΔR/R, the probe photon energy, and the pump-probe time delay,

respectively. For atomically thin WSe$_2$ on sapphire substrate, ΔR/R is proportional to absorption change. Non-resonant $\sigma_+$ pump of photon energy at 1.53 eV leads to a strong transient absorption signal for probes with the same polarization $\sigma_+$ (**b.**), but produces no transient response for probes with the opposite polarization $\sigma_-$ (**c.**). It demonstrates that non-resonant $\sigma_+$ pump significantly modifies A-exciton transition at the K valley, but not at the K' valley.

**Figure. 3 Time evolution and spectral lineshape of the valley-dependent transient reflection spectra for $\sigma_+$ pump and $\sigma_+$ probe configuration**. **a.** Time evolution of the pump-induced transient reflectivity ΔR/R at probe energies of 1.65 eV (blue line) and 1.7 eV (red line). Both signals reach the maximum magnitude at the pump-probe delay τ=0 with symmetric rise and decay dynamics (limited by our pulse width ~250 fs), which is characteristic of an instantaneous optical response. The transient reflectivity has opposite signs for 1.65 eV and 1.7 eV probes. **b.** The transient reflection spectrum at τ=0 (green circles). The transient reflectivity signal changes sign at the A-exciton resonance energy $E_A$=1.68 eV, and it shows a dispersive line shape. This transient reflection spectrum matches well with the derivative of linear absorption spectrum in monolayer WSe$_2$, and can be fully accounted for by a 4 meV blueshift of the K-valley A-exciton resonance (magenta line).

**Figure 4 Pump detuning and pump intensity dependence of the valley-selective optical Stark shift. a.** The transient reflection signal at 1.7 eV (left axis) and the corresponding optical Stark shift (right axis) as a function of the pump detuning energy (green dots). The dependence can be nicely described by an inverse proportional relationship (magenta line). **b.** The transient reflection signal at 1.7 eV (left axis) and the corresponding optical Stark shift (right axis) as a function of the pump laser intensity (green dots), which shows a linear scaling behavior (magenta line). The optical Stark shift can selectively shift the K-valley exciton transition by as much as 10 meV, which corresponds to a valley pseudomagnetic field of ~ 170 T.

**Acknowledgments:** This work was supported by Office of Basic Energy Science, Department of Energy under contract No. DE-SC0003949 (Early Career Award) and No. DE-AC02-05CH11231 (Materials Science Division). L.J.L. thanks the support from Academia Sinica and National Science Council Taiwan (NSC-102-2119-M-001-005-MY3). F.W. also acknowledges the support from a David and Lucile Packard fellowship.

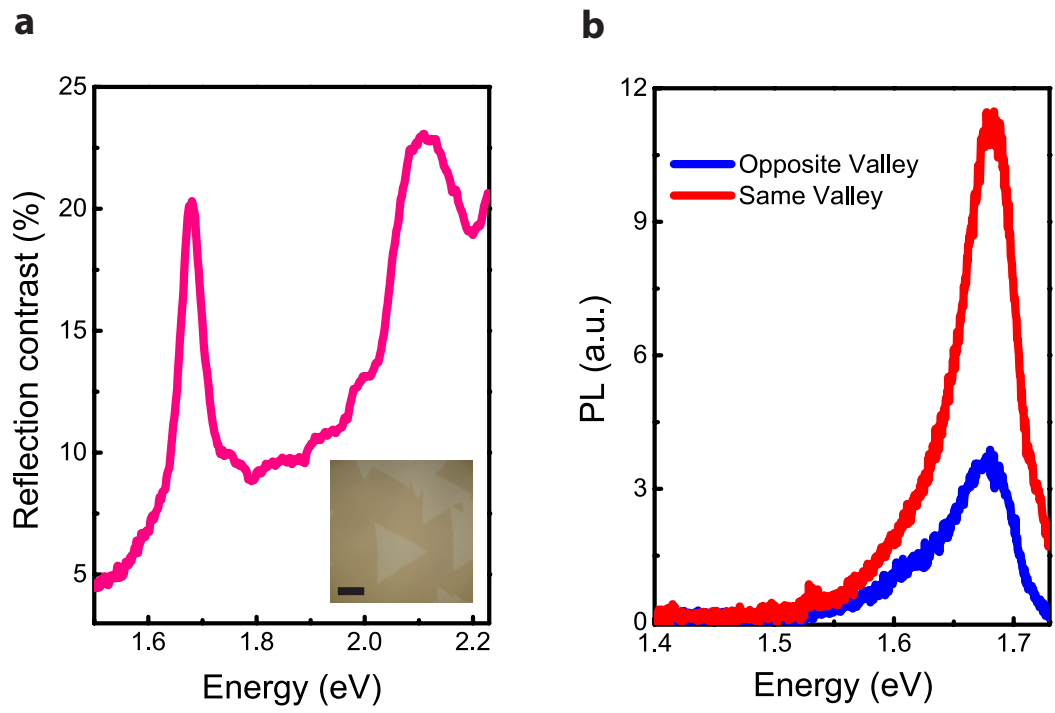

**Figure 1**

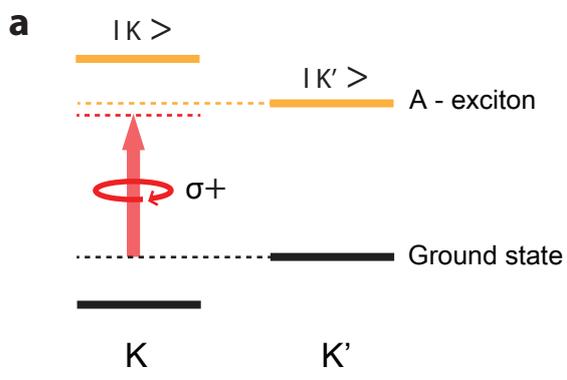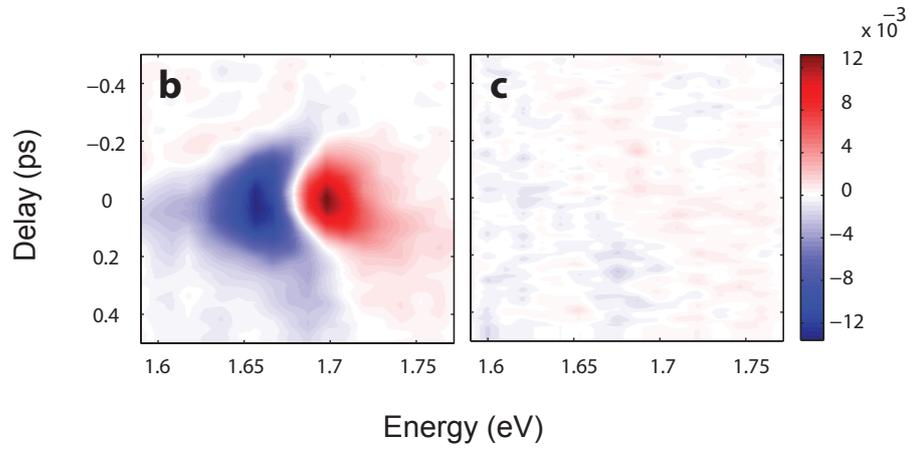

**Figure 2**

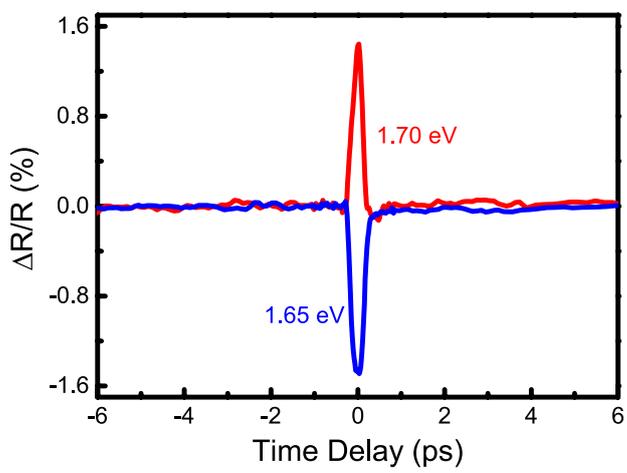 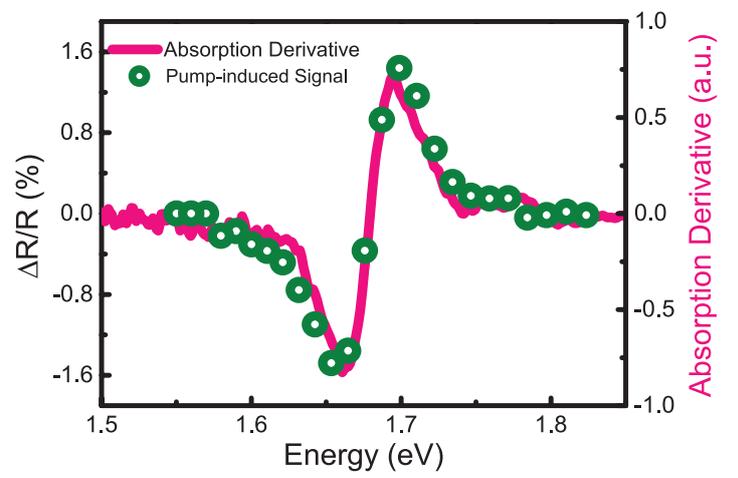

**Figure 3**

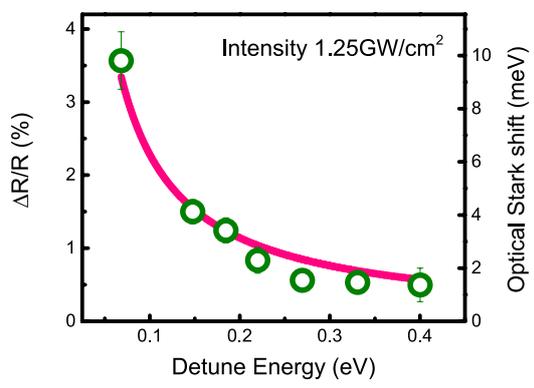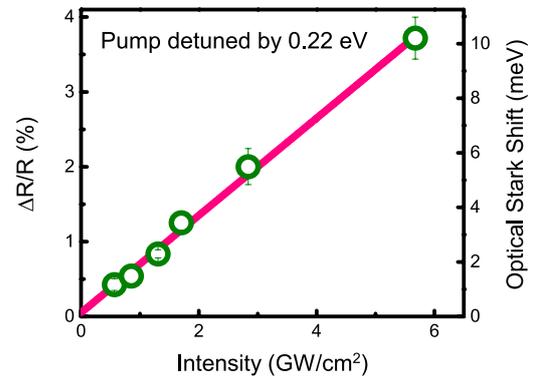

**Figure 4**